# Impact of street canyon morphology on heat and fluid flow: an experimental water tunnel study using simultaneous PIV-LIF technique

Yunpeng Xue [a,*], Yongling Zhao [b], Shuo-Jun Mei [c], Yuan Chao [d, e], Jan Carmeliet [b]

[a] *Future Resilient Systems, Singapore-ETH Centre, ETH Zurich, Singapore*
[b] *Department of Mechanical and Process Engineering, ETH Zürich, Zürich, Switzerland*
[c] *School of Atmospheric Sciences, Sun Yat-sen University, Zhuhai, China*
[d] *Department of Architecture, National University of Singapore, Singapore*
[e] *NUS Cities, National University of Singapore, Singapore*

**Abstract**

Urban areas are known for their complex atmospheric environments, with the building morphology having a significant impact on local climate patterns, air quality, and overall urban microclimate. Understanding the heat transport and fluid flow in complex urban environments is crucial for improving urban climate resilience, which remains an open frontier in the field of urban studies. To gain a more profound insight into the physical processes occurring in urban areas, particularly within street canyons, we conducted an experimental investigation in a large-scale water tunnel. This study involved the simultaneous examination of heat and flow fields, carried out at high spatial and temporal resolutions, utilizing Laser-induced Fluorescence (LIF) for heat analysis and Particle Image Velocimetry (PIV) for flow analysis. Our results of heat and flow in different street canyons indicate that the flow is significantly influenced by a combination of factors, including canyon configuration, the presence of buoyant force, and the magnitude of the approaching flow. The ventilation rate and heat flux from the street canyon, which are key factors shaping the urban microclimate, are found dominated significantly by the street canyon morphology. For instance, changing the aspect ratio of a street canyon results in a significant change of air ventilation rate, ranging from as low as 0.02 to as high as 1.5 under the same flow conditions. Additionally, canyons with high air ventilation rates exhibit significant heat flux removal at the canyon roof level, which is accurately described by the local Richardson number.

# 1 Introduction

As a result of multiple physical processes, Urban Heat Island (UHI), indicating the temperature difference between urban and rural neighbourhoods, is a complex phenomenon that varies across different spatial and temporal scales, ranging from microscale (a few meters) to macroscale (hundreds of kilometres). It has significant impacts on thermal comfort, public health, energy consumption, and sustainability of urban areas, and should be taken into careful consideration in urban planning and design [1-5]. To mitigate UHI, sustainable and economically sound urban design measures are being implemented, including the use of green infrastructures [6-9], new construction design and materials [10-13], and advanced energy-efficient technologies [14], among others. These measures require a good understanding of the airflow characteristics in urban areas, which play a crucial role in the dominant physical processes involving moisture and heat transport.

Street canyons are a ubiquitous element of urban landscapes, and understanding the flow behaviour around them, particularly the buoyant flow in an urban heat island environment, is a fundamental and critical aspect of urban climate research [15]. The temperature-driven buoyant force can significantly affect the flow structure, air ventilation, heat flux and temperature distribution in a street canyon, particularly under calm wind conditions [16-18]. Laboratory-scale experimental studies play an important role in advancing our understanding of the complex multi-physical processes in urban street canyons. By replicating controlled scenarios in a laboratory environment, typically in a fluid tunnel, researchers can isolate and measure specific variables and phenomena, providing valuable insights into the underlying mechanisms [19, 20] and critical validation information for numerical simulations [21, 22]. Non-invasive optical measurement techniques, such as laser doppler anemometer (LDA) [23, 24] and particle image velocimetry (PIV) [25-27] are commonly used to obtain detailed flow properties in street canyons without disturbing the flow. These techniques enable researchers to measure flow characteristics with high spatial and temporal resolutions, providing a comprehensive understanding of the flow behaviour in street canyons.

Other experiments have used simplified 2D street canyon models with heated surfaces placed in water tunnels to study the impact of canyon aspect ratio, surface temperature, and flow velocity on buoyant flow in the canyon [28, 29]. The flow characteristics under different canyon configurations and heating conditions were also studied using 2D street canyon models placed in 3D urban model arrays [25-27]. In one study, the impact of roof shape and building length on a 3-D buoyant flow in street canyons was studied in a wind tunnel [30], revealing the importance of considering full 3-D flow structures in street canyons. In shorter street canyons, where horseshoe vortex and corner vortex play more important roles in the canyon flow, thermal stratification resulted in different flow behaviours and pollutant dispersion [23, 24].

Understanding temperature profiles of the buoyant flow in street canyons is crucial for addressing a range of urban challenges, from mitigating the UHI effect and improving air quality to reducing energy consumption and associated greenhouse gas emissions. In these wind tunnel tests mentioned above, the temperature was measured by discrete thermal sensors [23, 24, 27], which only allows point measurements, with limited resolution and disturbance to the flow. Infrared thermography focusing on a black felt sheet that was mounted in the canyon also gave the temperature profile of the flow in the street canyon model [30]. The significantly induced disturbance to the flow by a black felt sheet and limited measurement accuracy were the main limitations. Hence, air temperature measurements were only used to qualitatively identify the impact of different street canyon configurations.

Understanding the intricate physical mechanisms at play within street canyons, especially those involving thermal-driven buoyancy flow, is a complex endeavour with far-reaching implications for the local microclimate. These implications encompass vital aspects such as air ventilation, pollutant dispersion, heat dissipation and removal. Despite ongoing research efforts, a comprehensive grasp of the interplay between heat transport and fluid flow in this context remains elusive. To enhance our comprehension of buoyant flow in street canyons, we created 3-D parametric urban models that closely resemble the actual urban morphology of Singapore (to be detailed in section 2.2), taking into account buoyancy effects. Through simultaneous PIV and LIF (Laser-induced fluorescence) measurements in a large closed-circuit water tunnel, we captured high-resolution heat and fluid flow behaviour in street canyons with different configurations. Our analysis of the heat and flow behaviours, air ventilation and heat flux characteristics under different conditions enabled us to identify the impacts of incoming flow, surface thermal conditions, and canyon configuration, which are discussed in detail in the following sections.

## 2 Experimental configurations
### 2.1 Experimental facilities

This study focuses on buoyant flows and utilizes a water tunnel setup to measure temperature and flow fields simultaneously via PIV-LIF measurements conducted in the ETH Zurich Atmospheric Boundary Layer Water Tunnel operated at Empa (Swiss Federal Laboratories for Materials Science and Technology). The water tunnel is equipped with a 110-kW pump capable of producing water speeds ranging from 0.02 to 1.5 m/s. It features a development section that measures 6-m long and a measurement section with a cross-sectional area of 0.6 × 1 m². Two sCMOS 16-bit dual frame cameras with a resolution of 2048 × 2048 pixels² at 25 Hz are included in the setup, which are aligned in the spanwise direction of the tunnel to focus on the same measurement plane.

In order to replicate realistic urban conditions, we employed nine stainless-steel plates, each measuring 330 mm × 330 mm, with conductive models placed on their surfaces. Beneath each plate, electrical heating elements were installed within sealed cavities to prevent direct contact with water. To ensure precise control and monitoring of temperature, we incorporated thermocouples in each plate to provide real-time temperature data. Our sophisticated control system facilitates both temperature monitoring and heating element regulation simultaneously. This capability allows us to consistently maintain the desired plate temperature throughout the experiment, ensuring accurate and reliable results. PIV measurements were obtained by employing a Litron 100 Hz Nd-YAG laser (532 nm) to excite the dye and illuminate the field-of-view. 10-micron hollow glass was used to seed the water flow with a particle density of approximately 40 to 60 in an integration window. To minimize laser intensity fluctuation caused by waves on the free air-water surface, we placed an optical boat above the water's free surface with its bottom submerged about 5 mm into the water. LIF images obtained from isothermal tests were used to determine the fluctuation in laser intensity, which amounted to 2.02%. To obtain temperature-dependent fluorescence, we utilized uranine due to the minor pulse-to-pulse intensity variation of the laser. Further details of its selection criteria can be found in [19]. As the emission peak of uranine is at the wavelength of 510 nm when the excitation is at 532 nm, a bandpass filter of 535-630 nm was adopted. The experimental setup in the water tunnel, including the laser, cameras, and building models, is presented in Figure 1.

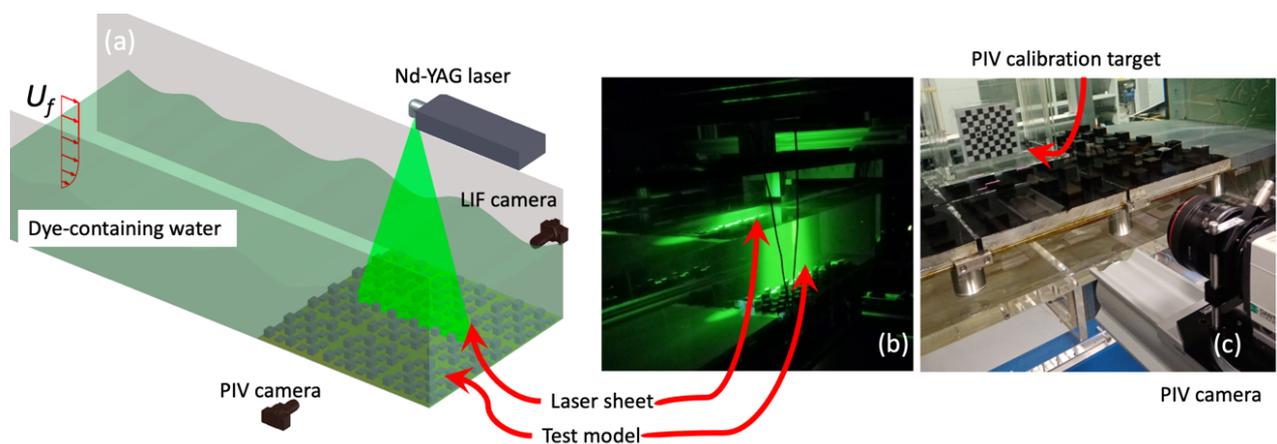

*Figure 1. (a) Water tunnel filled with dye-containing water and equipped with a PIV-LIF measurement system, (b) actual measurement process illuminated by the laser sheet, and (c) PIV camera and calibration target.*

## 2.2 Model configuration

To establish the parametric models, which are representative of Singapore scenarios with various urban densities, we calculated urban morphology density indices, i.e., site cover ratio ($\lambda_p$), average height ($H$) and sky view factor ($\Psi_{sky}$) in GIS platform with a resolution of 600 m × 600 m. Satellite data of the Singapore building geometry, which has been proved accurate and effective [5, 31], is used to calculate these parameters. The site cover ratio is the building footprint area normalized by the site area and the average height is the mean value of building heights (building footprint area-weighted). The sky view factor is calculated within a 100 m radius of the sky hemisphere based on the raster-based algorithm in 1 m × 1 m and aggregated into 600 m × 600 m. For urban models tested in this work, the site cover ratio, average height and sky view factor range between 0.21-0.27, 18-24 and 0.51-0.6, respectively, representing a typical low-density neighbourhood. Figure 2 illustrates a real urban area in Singapore alongside the laboratory-scale urban models employed in this study, complete with their respective configurations. It's important to note that the designed urban model and street canyon configuration have been scaled down from full-scale buildings at a ratio of 1:1000. Detailed information on the five different street canyons is given in Table 1, and visual representations of these configurations are provided in Figure 5. In case (a), the street canyon comprises buildings that have a height of 18 mm and a space between two buildings, or the width of the canyon itself, of 45 mm, resulting in an aspect ratio of 0.4. In case (b), the height of the buildings is increased from 18 to 24 mm, while the width remains the same as in case (a), i.e., 45 mm. In case (c) the width is reduced from 45 to 30 mm, while the height of the buildings remains the same as in case (b). In case (d) we have a step-up street canyon, where the first building has a height of 9 mm, while the second building is 27 mm, keeping the average height equal to 18 mm as in case (a). Case (e) is a step-down configuration, with heights of the first and second buildings equal to 27 and 9 mm. The width of the street canyon in cases (d) and (e) is equal to 45 mm as in case (a).

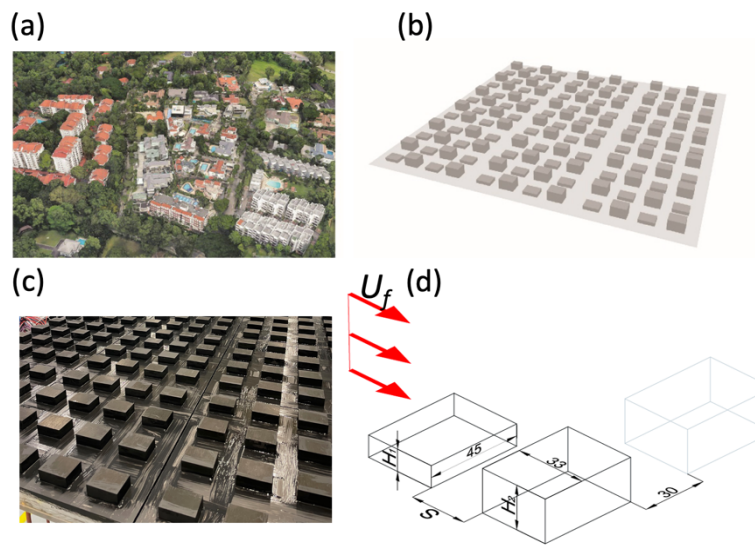

*Figure 2. (a) Typical low-density urban area in Singapore, (b) a designed urban model with variable height (case d), (c) stainless-steal models with the same height mounted on the heating plate (case (a)) and (d) detailed configuration of the tested street canyon (case (d)).*

*Table 1. Geometrical parameters of the tested street canyon models. $H_1$, $H_2$ and $H$ are the height of the building model and the average height of the canyon, $S$ denotes the spacing between the two building models, i.e., the width of the canyon, $A = H/S$ is the aspect ratio of the street canyon, Ri denotes the Richardson number at the lowest test velocity (0.03 m/s) using H, ΔT is the difference between the heating board surface and the freestream flow.*

| Case | $H_1$ (mm) | S (mm) | $H_2$ (mm) | H (mm) | A = H/S | ΔT (°C) | Ri | $\Psi_{sky}$ | $\lambda_p$ |
|---|---|---|---|---|---|---|---|---|---|
| a | 18 | 45 | 18 | 18 | 0.4 | ~20 | 0.8 | 0.6 | 0.21 |
| b | 24 | 45 | 24 | 24 | 0.53 | ~20 | 1.1 | 0.6 | 0.21 |
| c | 24 | 30 | 24 | 24 | 0.8 | ~20 | 1.1 | 0.51 | 0.27 |
| d | 9 | 45 | 27 | 18 | 0.4 | ~20 | 0.8 | 0.6 | 0.21 |
| e | 27 | 45 | 9 | 18 | 0.4 | ~20 | 0.8 | 0.6 | 0.21 |

## 2.3 Measurement methods and procedures

The water velocity in the tunnel in this study has been set at approximately 0.03, 0.06, and 0.15 m/s, with slight variations (< 2%) due to blockage of the urban model and water evaporation. These velocities are used to control the tunnel flow rate and determine flow parameters. The field-of-view (FOV) for the PIV-LIF measurements is positioned in the middle of the central street canyon, mounted on the central plate, covering the street canyon and the region above the building models to capture the in- and outflow from the street canyons. Velocity fields are obtained using PIV images, which are processed through cross-correlation with an integration window of 32 × 32 pixels² (equivalent to 10 pixels/1 mm). For each test case, the average velocity of the upper part of the PIV measurement (180 to 200 mm above the floor) is selected as the free stream velocity for further analysis.

Figure 3 illustrates the profiles of the normalized streamwise velocity distribution and turbulence intensity of the approaching flow in the isothermal and non-isothermal heating cases without building models on the heating plates. The velocity, in this case, is 0.03 m/s, and the temperature of the floor in isothermal and non-isothermal heating conditions is 22°C and 42°C, respectively, which are also the thermal conditions in the following tests. The boundary layer thickness ($\delta_{99}$) is 145 mm in the isothermal condition and increases to 175 mm when the floor is heated. A slight increase in turbulent intensity is also observed with a hot floor surface.

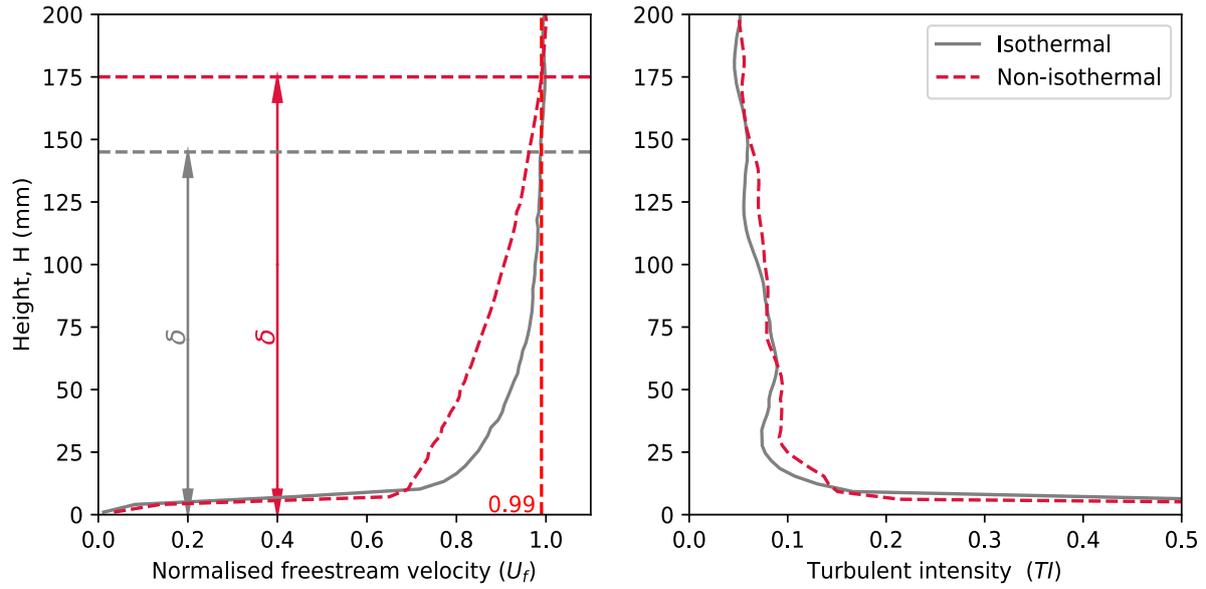

*Figure 3. Profiles of normalized freestream velocity and turbulence intensity of the approaching flow in the isothermal and non-isothermal heating cases without the building models on the heating plates. The velocity is 0.03 m/s. (a) normalized freestream velocity and (b) turbulence intensity.*

Temperature profiles are obtained by post-processing LIF images. The local intensity of the LIF image depends on the uranine concentration, laser intensity, and fluid temperature. To ensure a good linear relationship between the measured local intensity and the fluid temperature in a flow with uniformly mixed uranine, stable laser power and a constant uranine concentration are maintained during each measurement. The relationship is calibrated daily to obtain reliable measurement results, considering the decreasing exciting level of the dye with time. A detailed description of the calibration process is found in [19]. However, the transparent box used in the calibration may introduce some measurement uncertainty as it adds an extra layer of transparent Perspex, which does not exist in actual measurements. To achieve good measurement accuracy, the uranine concentration in this study is maintained at about 2 mg/L. As the laser beam is spread to a sheet, a spatial correction of the laser intensity distribution is required for each non-isothermal case. This is achieved by using a laser intensity profile obtained under isothermal conditions. Additionally, several thermocouples are used to measure the ground and fluid temperatures, which are then used to validate and correct the temperature measurements.

The simultaneous measurement of velocity and temperature is achieved by overlaying the two fields from the same FOV. For each test configuration, 1500 pairs of images are captured at a frequency of 15 Hz, resulting in a recording time ($T$) of 100 seconds for statistical analysis. The uncertainty of the velocity fields is estimated to be $10^{-5}$ m/s using integration window-based statistics with sub-pixel accuracy at 1/10 pixel, which is two orders of magnitude smaller than the freestream velocity. The temperature measurement uncertainty is dependent on the uranine emission spectrum, optical setup, and pulse-to-pulse laser intensity variation. The intensity-to-temperature ratio is approximately 450, resulting in an instantaneous temperature field uncertainty of 0.002°C. For time-averaged statistics, the pulse-to-pulse laser intensity variation (2.02%) yields an uncertainty of approximately 0.09°C.

When describing non-isothermal flow, the bulk Richardson number is used to express the ratio of the buoyancy term to the flow shear term as [19]:

$$Ri = \frac{g\beta\Delta T H}{U_f^2} \tag{1}$$

where $\Delta T$ is the temperature difference between the floor and the freestream water, $\beta$ denotes the thermal expansion coefficient of water, $g$ is the acceleration due to gravity, $H$ represents the height of the downstream building model, and $U_f$ is the freestream velocity. The selection of our experimental parameters was guided by the scaling of the Richardson number. Taking these considerations into account, we opted for a representative urban climate scenario featuring a wind speed of 2.7 m/s and a temperature difference of 10 degrees Celsius between the surface and the ambient air. Consequently, we set the water velocity and surface temperature at 0.03 m/s and 42 degrees Celsius, respectively, to maintain an equivalent Richardson number across the experiments. Table 1 provides an overview of the bulk Richardson numbers for various canyon configurations, which will be further discussed in Section 7.

## 3   Characteristics of the canyon flow
### 3.1   Example of the Instantaneous flow and temperature fields

Figure 4 shows an example of the instantaneous velocity and temperature measurements in case (c), with $U_f$ = 0.03 m/s and $Ri \approx$ 1.1. Distinct thermal plumes can be seen developing and shedding from the heated ground and building surfaces, resulting in a significant increase of 5 ~ 6°C in water temperature within the street canyons. Some of the accumulated heat is carried away by the updraft and subsequently flushed downstream by the approaching flow above the rooftop. The flow is clearly dominated by buoyancy, as evidenced by the strong upward flow ascending to three times the building height. The standing-vortex typical for a forced convection case is not seen in the canyons as upward plumes alter the flow structure, as discussed in detail in [26]. The canyon flow focused on is highlighted in the boxed area.

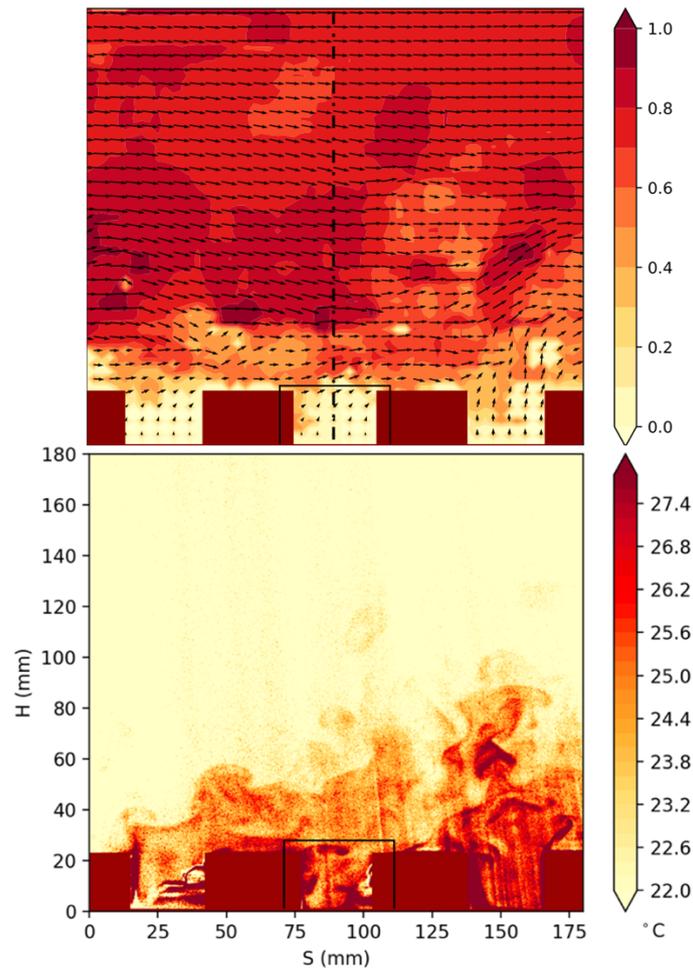

*Figure 4. Normalized instantaneous velocity field (top) and corresponding temperature field (bottom) across the street canyons in the test case (c) with free stream velocity of 0.03 m/s and floor temperature of about 42 degrees.*

## 3.2 Impacts of the canyon configuration

### 3.2.1 Temporal-averaged velocity field

Figure 5 provides insight into the impact of the street canyon configuration on the flow pattern, by summarizing the time-averaged velocity fields normalized by the free stream velocity in different street canyon configurations. For instance, $a_1$ and $a_2$ depict the flow pattern in canyon case (a) under isothermal conditions (subscript 1) and with the floor surface temperature at 42°C (subscript 2), respectively. In the isothermal tests, the velocity fields in case $a_1$, $b_1$ and $c_1$ reveal the presence of a canyon vortex, with aspect ratios of 0.4, 0.53, and 0.8 as listed in Table 1, respectively. As the canyon aspect ratio increases, the vortices become stronger, as evident from the out-of-plane vorticity depicted in Figure 8. This phenomenon of a strengthening canyon vortex in street canyon models with a higher aspect ratio, and the generation of a double vortex when the aspect ratio exceeds 2, has also been observed in other experimental [32, 33] and numerical [34-36] studies.

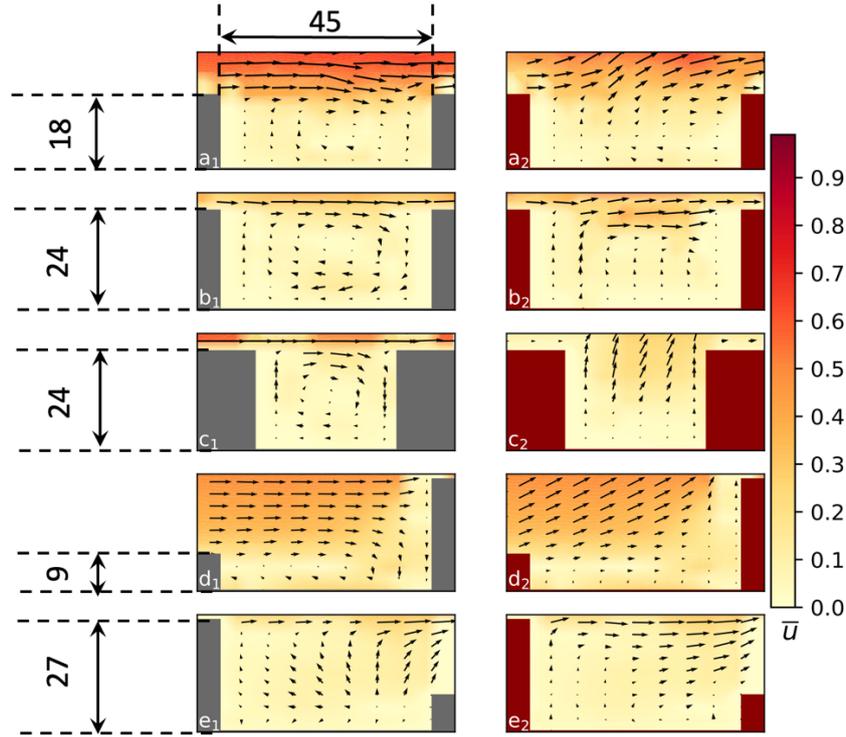

*Figure 5. Time-averaged velocity vector fields and freestream velocity magnitude fields for different configurations are normalized by freestream velocities, respectively. The subscript 1 & 2 denote the isothermal and non-isothermal conditions.*

In canyons with variable building heights, the flow patterns agree well with those observed in previous experiments [25, 30]. In the isothermal test of the canyon (d), the higher downstream building model induces a downward flow flushing along its windward surface, and the lower height of the upstream model limits the formation of a canyon vortex. In the street canyon with variable model heights, a canyon vortex can still be observed when the aspect ratio of the upstream model is not too small [25, 30, 36]. In non-isothermal conditions, the canyon flow becomes more buoyant due to convective heating from the heated ground surface, which suppresses the formation of the canyon vortex and leads to significant updraft flow, particularly in cases $a_2$ and $c_2$. The thermal plumes depicted in Fig. 4 and the updraft flow in heated canyons (Fig 5 $a_2$, $c_2$ & $e_2$) highlight the transport heat flux process from the street canyon.

Figure 6 presents the streamwise velocities along the canyon centreline (as shown in Fig. 4) for the tested cases with a freestream velocity of 0.03 m/s and a plate temperature of 42°C. The velocity profiles within the canyons vary, with some cases showing negligible streamwise velocity (such as in canyons ($b_2$) and ($c_2$)), and negative velocity near the ground in the canyon ($a_2$). However, there is a good similarity in the velocity profiles over the building models (H > 27 mm).

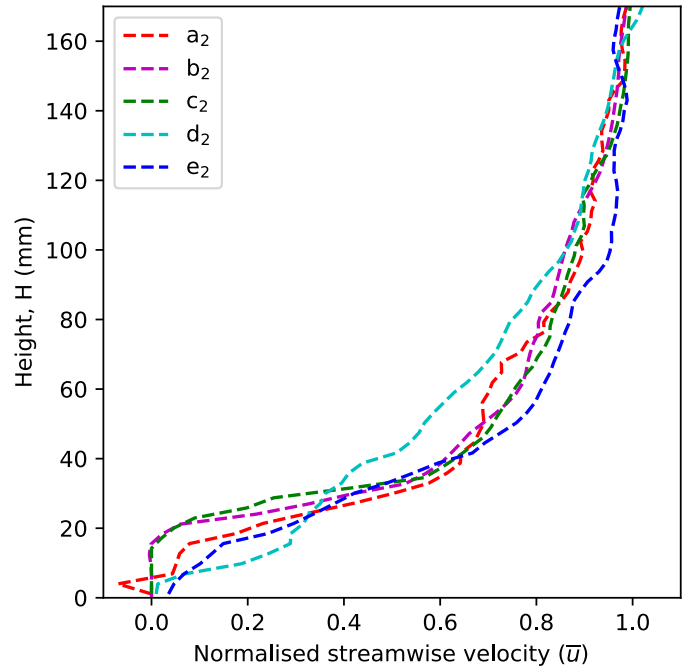

*Figure 6. Normalized streamwise velocity along the canyon centreline in different tests with a freestream velocity of 0.03 m/s.*

The normalized average vertical velocity profiles ($\bar{v}$) presented in Figure 7 emphasize the updraft flow in the non-isothermal condition. In the isothermal cases ($a_1$, $b_1$ & $c_1$), the downwards and upwards flow indicated by the blue and red contours show the existence of the canyon vortex and the significant part of the downwards flow in $d_1$ is caused by the blockage of the higher downstream building model. In the non-isothermal conditions, strong upward flow is observed in case $a_2$, $c_2$ and $d_2$, which is dominated by the buoyance force from the heated surfaces. While, the negligible enhancement of the vertical flow in case $b_2$, could be caused by the suppression of the approaching flow at the canyon opening. This demonstrates the significant influence of the canyon configuration, particularly the aspect ratio, on the flow behaviour.

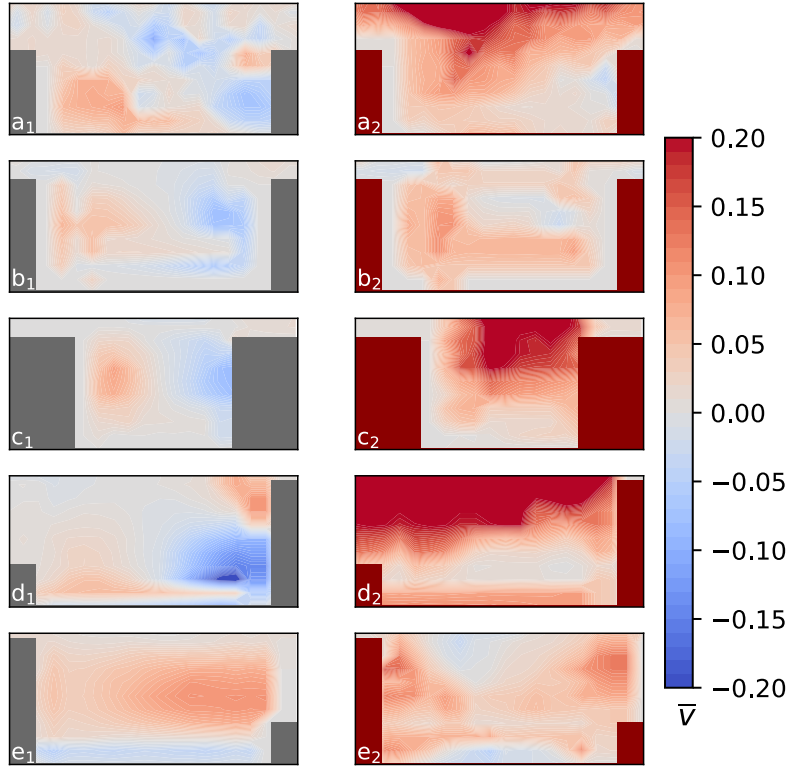

*Figure 7. Time-averaged vertical velocity profiles for different configurations are normalized by free stream velocities, respectively.*

### 3.2.2 Vorticity and turbulent intensity

Out-of-plane vorticity profiles ($\overline{\omega_z}$) in different tests are averaged, normalized and presented in Figure 8. The presence of canyon vortices in the isothermal cases is evident in the vorticity field, with stronger vortices observed in $b_1$ and $c_1$. However, in the heating cases $b_2$ and $c_2$, the coherent vortex in the canyon is disrupted by the thermal buoyant flows from the heated surfaces and the plume from the heated horizontal ground. Furthermore, the freestream turbulent intensity (*TI*) in the canyons is calculated using the time series of measurement data, and the *TI* values in case $a_1$, and $a_2$, and the ratio of the non-isothermal case over the isothermal case are presented in Figure 9. A higher turbulent intensity is observed in the central and over the canyon in both isothermal and heating conditions. The lower *TI* near the building model surfaces is due to the lower velocity fluctuation magnitude near the surface. The *TI* ratio demonstrates a significant increase in fluctuation in the canyon for the non-isothermal case, especially near the heating surfaces and ground. The heating source has a greater impact on the flow in the nearby region, resulting in a higher increase in the ratio near the surfaces. It is worth noting that all five test cases exhibit similar impacts of the heating surfaces on the turbulent characteristics, and thus, only one case is presented here.

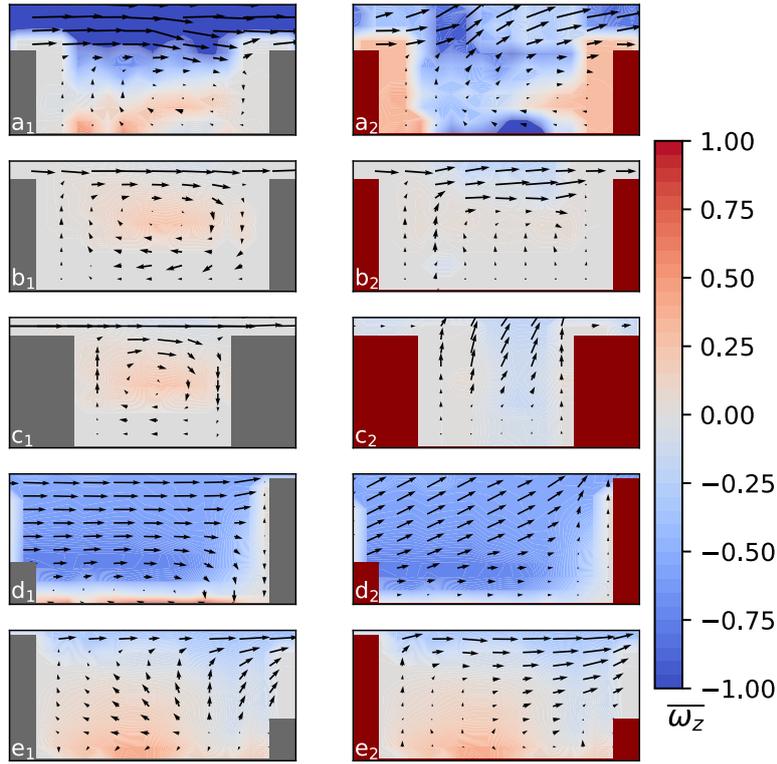

Figure 8. Time-averaged out-of-plane vorticity in different configurations with the averaged velocity field shown by the vectors.

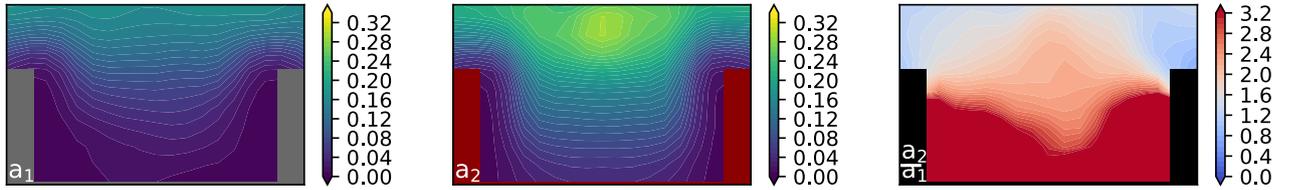

Figure 9. Turbulent intensity profiles in the canyon for case (a): isothermal (left), non-isothermal (middle) and increase of the TI calculated by the ratio of non-isothermal over the isothermal case (right), respectively.

### 3.2.3 Quadrant analysis

Quadrant analysis is a useful tool for identifying the dominant modes of momentum transport and distinguishing between different types of turbulent transport processes in street canyon flow. By examining the streamwise ($u'$) and vertical ($v'$) fluctuations of the velocity components, the momentum flux ($u' \times v'$) can be characterized as an outward interaction ($u' > 0$, $v' > 0$), ejection ($u' < 0$, $v' > 0$), inward interaction ($u' < 0$, $v' < 0$), or sweep ($u' > 0$, $v' < 0$). Previous studies have used quadrant analysis to classify the different events occurring in the street canyon region [25, 26, 37]. In this study, small magnitudes of the inward and outward interactions suggest that the dominant events in all cases are the ejections and sweeps, particularly in case $c_2$ and $d_2$. Hence, time-series data of the ejections (shown in red) and sweeps (shown in blue) at the roof level, as indicated by the red dashed line, were normalized by the freestream velocity ($U_f^2$) and plotted in Figure 10. The strong ejections and sweeps in $c_2$ and $d_2$, as well as their increases due to the heated surfaces, are depicted in the spatiotemporal plot at the canyon roof level. The periodic appearance of the sweeps and ejections, where high ejections are followed by high sweeps, indicates an intermittent behaviour and suggests the presence of coherent structures, such as thermal plumes.

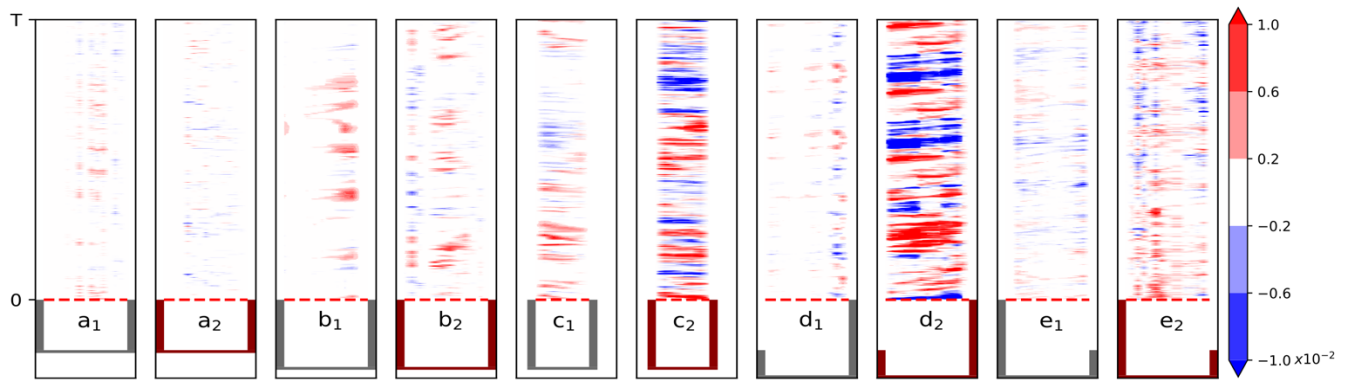

Figure 10. Time series of ejections (red) and sweeps (blue) at the canyon roof level in different tests.

### 3.3 Impacts of the freestream velocity

The impact of freestream velocity on the flow mechanism was studied by increasing it from 0.03m/s to 0.06 and 0.15 m/s. In test canyon (b), Figure 11 shows the velocity fields for different freestream velocities and heating conditions. A vortex-dominating flow is observed in the canyon under isothermal conditions. With an increase in freestream velocity, the vortex is further pushed downstream closer to the windward building. When the ground and building models are heated ($Ri = 1.1$), the buoyant flow becomes dominating in the canyon. However, the approaching flow still has a significant suppression effect on the buoyant flow in the current condition, as indicated by the negligible increase in the vertical velocity (also shown in Figure 7, $b_2$). As we further increase the freestream velocity, or in other words, decrease the Richardson number by increasing the velocity component, the impact of the heating source becomes weaker, which is indicated by the less change in flow pattern. At a Richardson number of 0.044, when the shear force dominates the flow, a canyon vortex is observed. Similar results are captured in case (a) and (c). For example, the flow pattern in case (c) is shown in Figure 12. Due to the smaller canyon width, a higher resolution of the velocity vectors is used here to avoid missing any important information. The isothermal tests indicate a similar canyon vortex as in case (a) and its shift to a downstream position with an increase in velocity. On the same heating condition, the smaller canyon leads to a more significant buoyant flow, as shown by a large amount of updraft flow in case $c_2$. Again, at the Richardson number of 0.044, when the buoyant flow has negligible influence, a canyon vortex is observed.

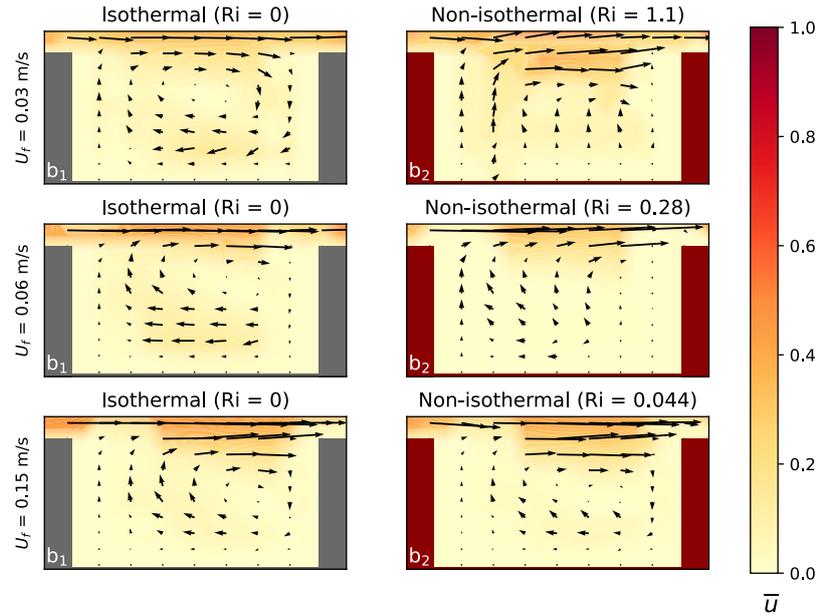

*Figure 11. The normalized velocity fields in canyon (b) at different freestream velocities and heating conditions.*

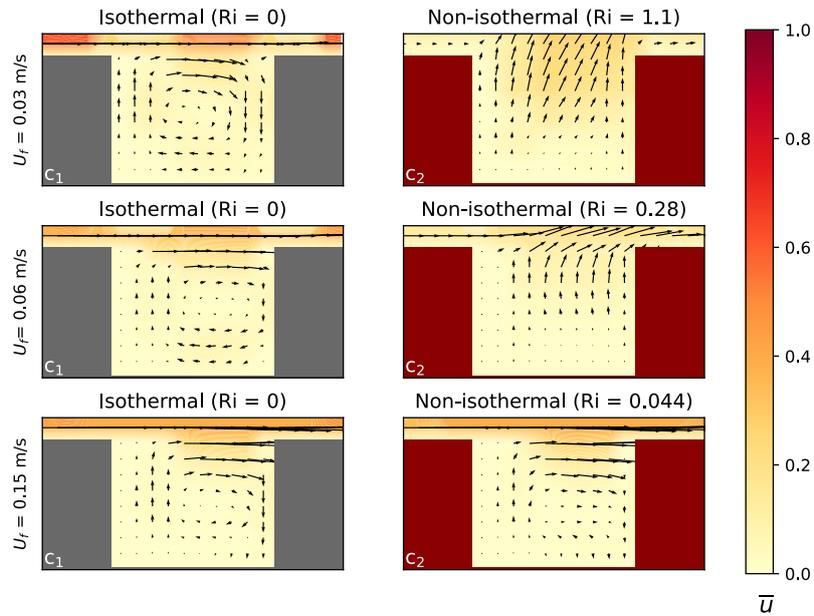

*Figure 12. The normalized velocity fields in canyon (c) at different freestream velocities and heating conditions.*

To study the flow in street canyons with variable heights (case (d)), we present in Figure 13 the flow under different flow and thermal conditions. The impact of heating the ground and building surfaces on flow behaviour is evident from the significant buoyant upward flow at a Richardson number of 0.4. As the Richardson number decreases, the effect of buoyancy weakens, leading to less pronounced changes in flow patterns under isothermal and non-isothermal conditions. These observations agree with previous studies [25, 27] on the impact of Richardson number on flow in street canyons with variable heights.

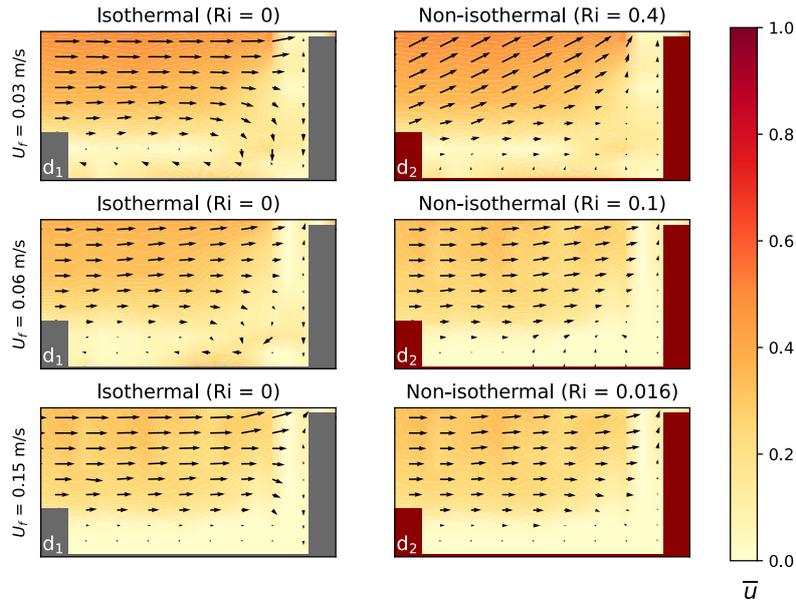

*Figure 13. The normalized velocity fields in canyon (d) at different freestream velocities and heating conditions.*

# 4 Temperature profiles in the canyons

## 4.1 Temporal-averaged temperature field

One of the advantages of the current PIV-LIF measurement is the high resolution and accuracy of the temperature measurements without disturbing the flow. Temporally-averaged temperature distributions are obtained to understand the physical processes and perform further analysis of the heat flux. For example, Figure 14 shows the temperature and velocity fields in case ($a_2$), allowing the visualization of the heat and fluid flow. The temperature gradient generated by the heat sources, as well as the resulting heat plumes and updraft flow, are clearly shown in the left figure. An expanded view of the heat and fluid flow in the canyon region is presented on the right. The temperature and velocity fields have good consistency, such as the heat plume from the left building model and the vortex structure in the street canyon. In the non-isothermal case, a relatively low-temperature region near the left bottom corner can be observed which might be explained by the 3D street canyon flow, where a flow at a lower temperature enters the region from the two sides of the canyon, and the horseshoe vortex generated from the approaching flow.

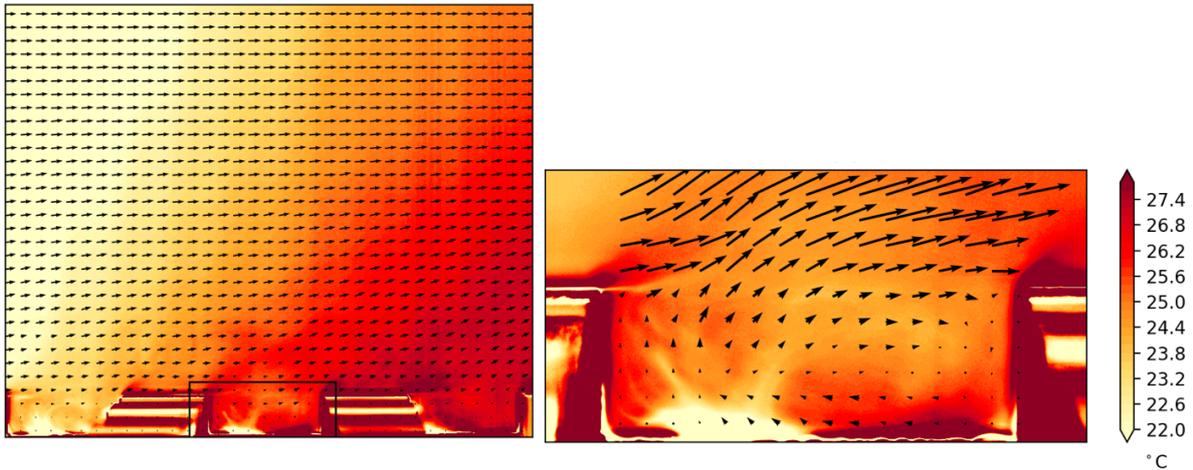

*Figure 14. left: FOV of the temporal-averaged temperature and velocity fields in the test case (a), right: Enlarged detail of the temperature and velocity profiles in the focused street canyon.*

## 4.2 Impacts of the canyon configuration

We plot the vertical temperature distributions along the centreline in different canyons, as well as the temperature profile in the turbulent boundary layer (TBL), in Figure 15. The red dashed line represents the centreline of the canyon, along which the fluid temperature is plotted, the grey dashed line indicates the roof level of the canyon and the vertical black dash line at 22 degrees indicates the ambient temperature of the flow. It is noticed from Figure 14 that at the bottom of the canyon, the strong reflection of the laser sheet (the light and dark regions near the floor) limits the measurement accuracy close to the ground surface. Therefore, the effective temperature profile is selected from 3 mm above the canyon bottom in the current work.

Under the same thermal condition, the temperature profile in the canyon is directly determined by the updraft buoyant flow and shear layer flow at the canyon opening, which are dominated by the canyon configuration and the approaching flow. It is interesting to observe a zigzag shape of the temperature distribution in the canyons, which is attributed to the 3D street canyon flow pattern. The first decreasing tendency of the temperature from the ground shows the existence of a thermal boundary layer and the impact of the entering flow from the canyon's two sides. The heat flux from the building models increases the temperature along the height until it reaches another maximum above the building models. The thermal plumes from canyon walls and model roof merge above the roof level, contributing to the local high temperature as evident in Fig. 14.

Canyon configurations (a), (b), and (c) share identical model heights but exhibit varying aspect ratios. When comparing Canyon (a) and (b), which both feature taller building models, it becomes apparent that the influence of buoyant flow diminishes in comparison to the shear layer flow. This reduced impact is evidenced by the observed vortex-type flow in the canyon (refer to Fig. 5). Consequently, this alteration leads to lower temperatures along the centreline of case (b), suggesting that the heated flow escapes from the sides of the canyon rather than through the canyon roof opening, a phenomenon supported by the ventilation rate data discussed in Section 5.1. In the case of the canyon (c), a slightly different temperature distribution prevails, primarily due to its larger aspect ratio. This larger aspect ratio enhances buoyant flow and reduces the suppression of the incoming flow. Consequently, under similar thermal conditions, more heat is transferred from the heating board to the fluid through the conductive models. As a result, the temperature profile in case (c) exhibits a slower decrease in temperature along its height. Canyons (d) and (e), which are derived from the same urban model, exhibit identical temperature profiles above the canyon roof level but starkly contrasting profiles within the canyon. The distinctive flow patterns within the canyon account for the variations in centreline temperature. Additionally, we have included the temperature profile in the turbulent boundary layer (TBL) without mounted building models to facilitate a clear comparison. This profile highlights the effect of conductive building models in carrying heat from the heating board and releasing it to the fluid at higher elevations, causing an upward shift in the temperature profile as illustrated in Figure 16.

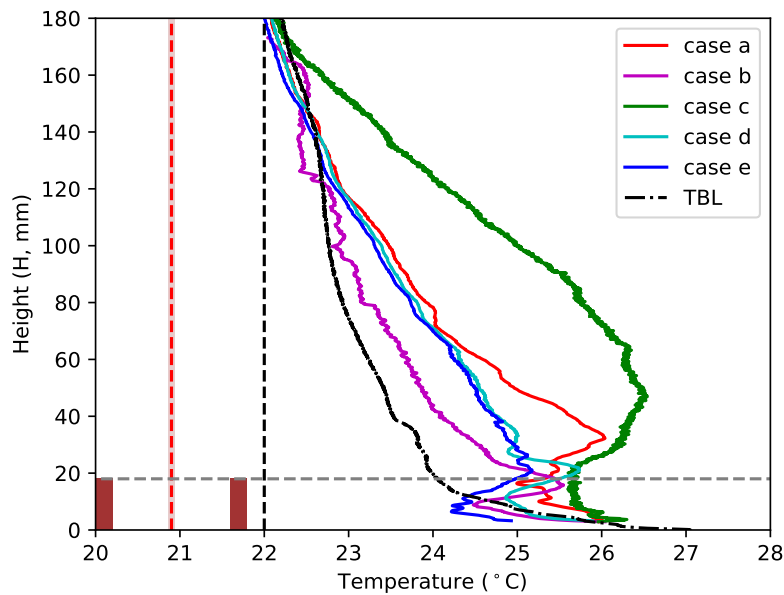

*Figure 15. Vertical temperature distributions along the canyon centreline in different cases with a comparison with the turbulent boundary layer temperature profile. The red dashed line represents the canyon centreline, along which the fluid temperature is plotted, the grey dashed line indicates the roof level of the canyon and the vertical black dash line at 22 degrees indicates the ambient temperature of the flow.*

Comparing the smooth temperature distribution of the BTL, the different profiles in the canyons, particularly below the roof level, are the results of the complex flow pattern within the street canyon. These canyon centreline temperature distributions have good agreement with the computed temperature profile at city scale [36], including the increasing temperature along the height from the ground to the maximum temperature above the building roof level and then decreasing tendency in the further higher region until reaching the freestream temperature. The main difference is that the thermal boundary layer is not seen in the large-scale simulation. As the thickness of the thermal boundary layer is always negligible when compared to the scale in the simulation, this also explains the negligible influence of the thermal boundary layers on the dominating flow structure in canyons and hence on the actual urban climate [22, 38, 39].

## 4.3   Impacts of the freestream velocity

Figure 16 presents the vertical temperature distributions in case ($a_2$) at three freestream velocities. It is shown that the temperature along the canyon centreline decreases with an increase in the freestream velocity. As heat is carried upwards by the buoyant flow and flushed away by approaching flow, a flow at a higher velocity has a greater heat removal capability and hence results in a lower temperature. The temperature generally decreases from the ground until reaching the steady freestream temperature at a certain height, which decreases from about 180 mm above the ground at a freestream velocity of 0.03 m/s to around 60 mm at 0.15 m/s. A zigzag distribution of the centreline temperature is observed at low velocity. Mixing with the approaching freestream flow, the flow temperature drops from about 40 mm above the ground at 26.1 °C to the freestream temperature.

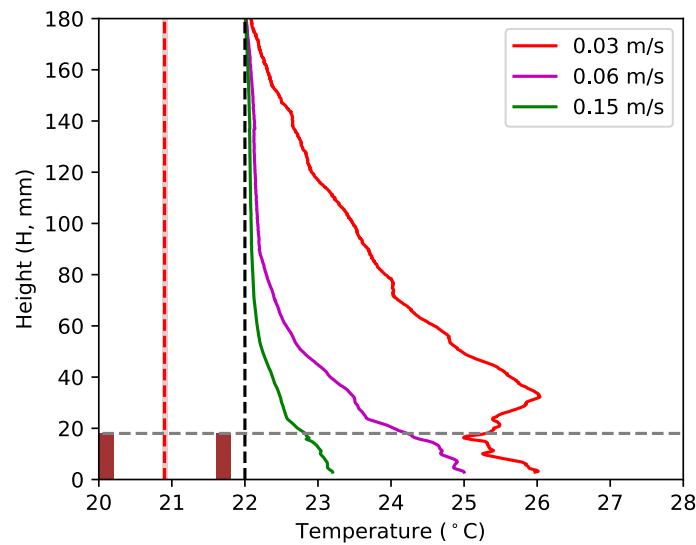

*Figure 16. The temperature along the centreline of the canyon (a) at different velocities.*

Vertical temperature distributions at different locations within the canyon are shown in Figure 17. On the left side of the street canyon, the temperature decreases noticeably from the roof level due to mixing with the approaching flow, and the decrease is more significant at higher velocities. As the flow passes over the canyon opening from left to right, the maximum temperature is also pushed higher at the right side by upwards buoyant flow. A thickened thermal boundary layer along the flow direction can also be observed in the figure. Part of the approaching flow enters the canyon from the right wall, leading to a thicker thermal layer on the right side. This portion of the flow moves reversely towards the left near the ground, continuously absorbing heat from the ground and mixing with the flow entering from the canyon side, resulting in a thinner thermal layer and lower temperature near the ground on the left side.

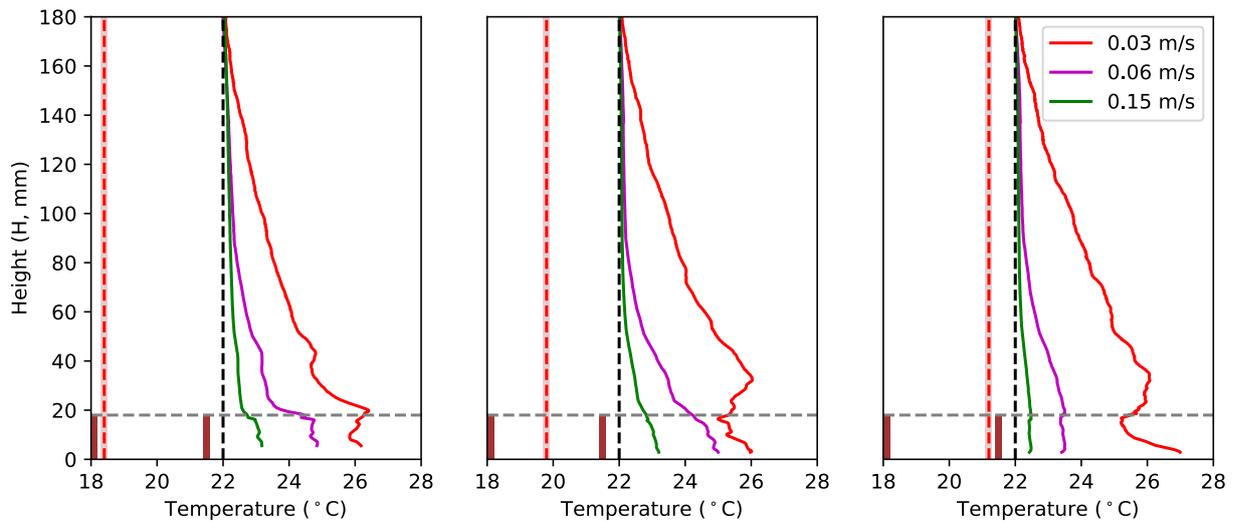

Figure 17. Vertical temperature distribution at different locations of the test case (a).

## 5 Roof-level ventilation

Thermal buoyancy is important for urban natural ventilation, especially during heat waves with high UHI intensity and weak wind conditions. To evaluate its importance, the canyon volumetric ventilation rate, $Q'$ is calculated based on the instantaneous vertical velocity at the roof level ($v$), given as:

$$Q' = \frac{\tau}{V_c} \int_A v \, dA \qquad (2)$$

where $\tau$ denotes the reference time and is calculated as $\tau = 2(H + S)/(2U_f/3)$, $V_C$ represents the unit volume of the canyon, and $A$ is the ventilation area of interest.

### 5.1 Impacts of the non-isothermal heating

The spatial and temporal ventilation rates at the canyon roof level are presented in Figure 18. The positive ventilation (red) indicates flow entering the canyon from the two sides and exiting from the roof level. Negative (blue) values represent flow entering the canyon from its rooftop opening and exiting from the sides. It can be concluded from the figure that with a higher ground temperature, the ventilation rate over the street canyon increases in most of the street canyons, such as cases (a), (c), and (d). Alternating positive (red) and negative (blue) ventilation rates in configurations ($b_2$ & $d_1$) are mainly due to an inward flow eventually combined with a canyon vortex. The fluctuating pattern also implies unsteady flow behaviour and the potential existence of periodic flow structures, such as the plumes, as indicated by the shear stress presented in Figure 10.

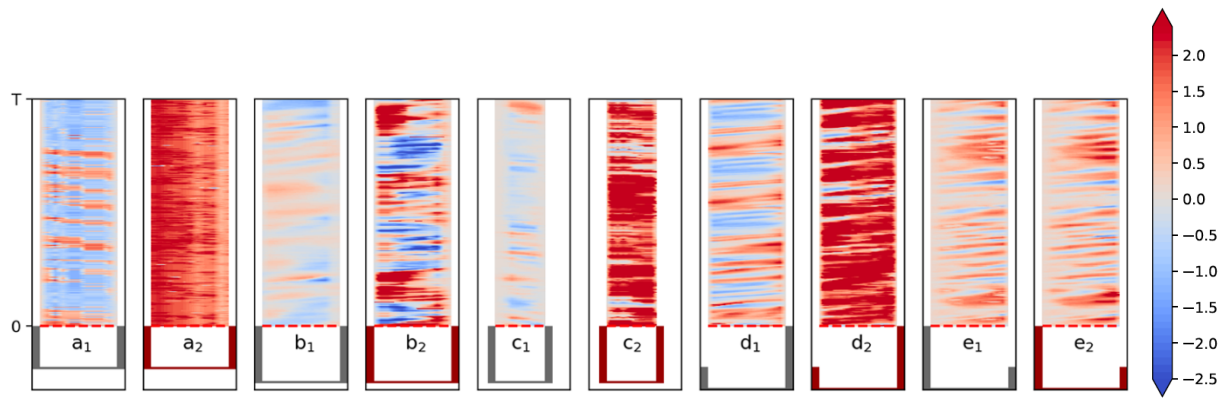

Figure 18. Spatiotemporal plot of the ventilation rates at the canyon roof level for the different cases.

Integrating the ventilation rate over the canyon width, Figure 19 shows the time series of the canyon-wise ventilation rates. The grey and red lines represent the fluctuating ventilation rate for isothermal and non-isothermal heating cases, respectively, and the dashed lines indicate the averaged ventilation rate with values given on the right side. Strong fluctuations in heating situations are observed in canyons (b), (c), and (d), indicating the unsteady updraft flow caused by buoyancy force or heat plumes. It is reasonable to conclude that in isothermal conditions, the flow exchange from the canyon to ambient flow is low, as indicated by the small magnitude ventilation rate, which is only determined by the canyon configuration. When the ground and building surfaces are kept at a high temperature, the buoyancy force tends to produce an updraft flow, leading to a significant increase in ventilation rate, as observed in cases (a), (c), and (d). For instance, in the street canyon case (a) when the approaching flow is set at 0.03 m/s, the ventilation rate increases from -0.353 in the isothermal condition to 0.91 at a temperature of around 42 °C. However, the increase in the ventilation rate is not significant in canyons (b) and (e) due to thermal stratification, which means the shear layer flow over the canyon opening dominates the flow and suppresses the updraft flow.

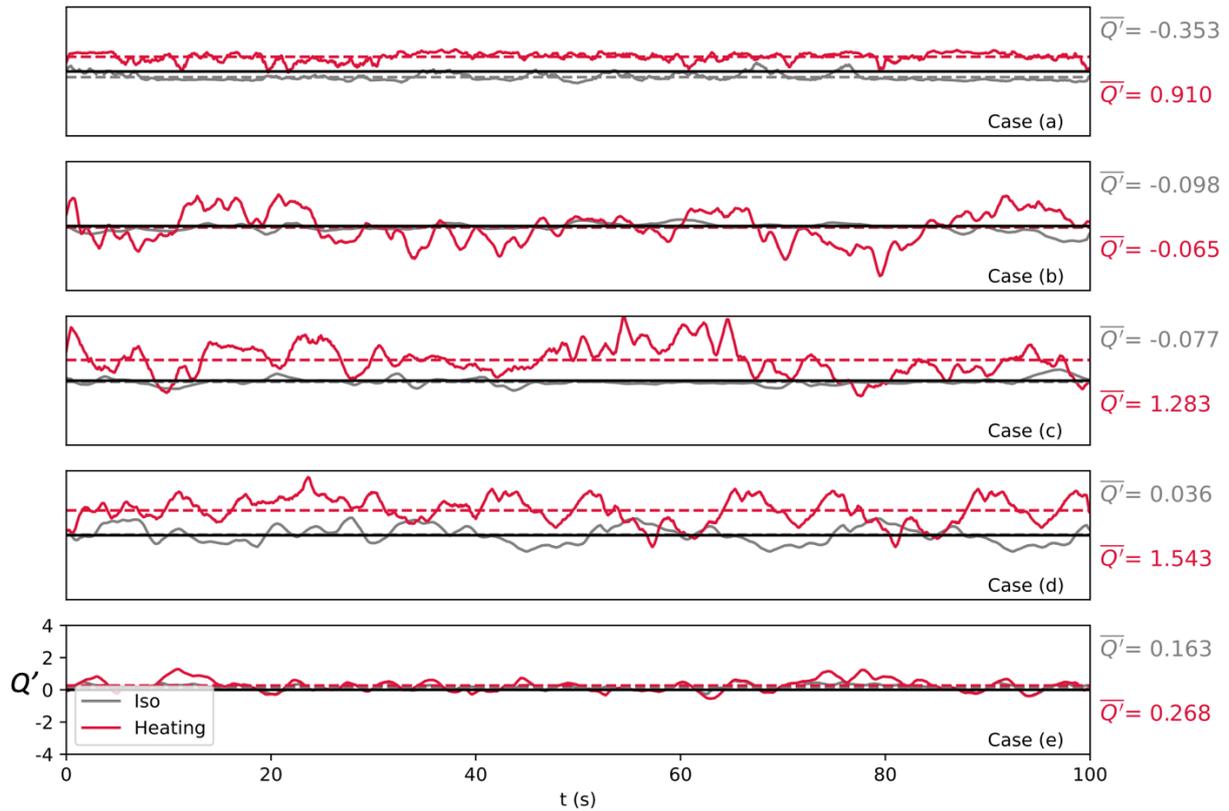

*Figure 19. Time series of the ventilation rate over the canyon roof level on isothermal (grey) and non-isothermal heating (red) conditions with the averaged values given on the right.*

## 5.2 Impacts of canyon configuration and freestream velocity

Figure 20 summarizes the ventilation rates for different flow, thermal, and canyon conditions. The labels $U_1$, $U_2$ & $U_3$ refer to tunnel velocities of 0.03, 0.06, and 0.15 m/s, respectively. As the velocity increases, the corresponding Richardson number decreases to 1/4 and 1/25 of the values listed in Table 1. It is immediately apparent that canyons (a), (c), and (d) exhibit large ventilation rates under non-isothermal conditions with a tunnel velocity of 0.03 m/s. For higher tunnel velocities ($U_2$ & $U_3$), smaller Richardson numbers indicate weaker impacts of the buoyancy flow. Plumes from the ground at smaller Richardson numbers cannot penetrate the shear layer flow, resulting in negligible ventilation enhancement in almost all tested cases. At low velocity ($U_1$), canyons (a), (c), and (d) exhibit significant ventilation enhancement (1.26, 1.36 and 1.51, respectively) due to thermal enhancement, and at a tunnel velocity of 0.06 m/s, canyon (c) still exhibits an enhancement of 0.28. For other conditions, negligible impacts of buoyant flow on ventilation rate are observed.

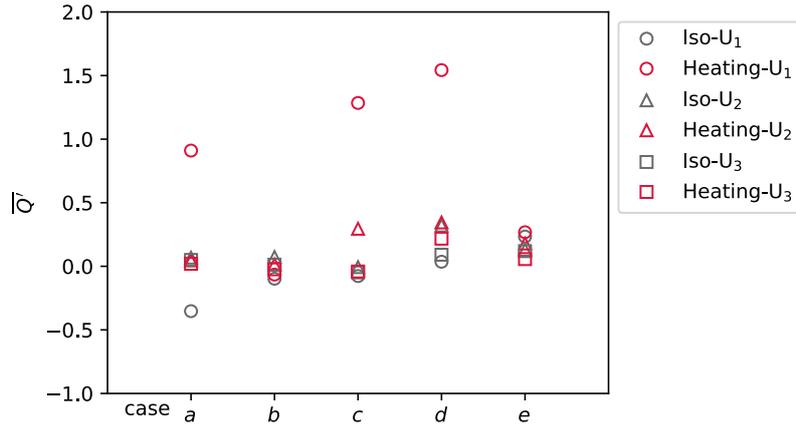

*Figure 20. Averaged ventilation rates in different street canyons (a to e) with variable velocity $U_i$ and heating conditions.*

# 6   Heat removal performance in different street canyons

Buoyant flow can significantly contribute to heat removal from the canyon by providing additional upward momentum and inducing a vertical heat flux from the ground to the free stream flow. With simultaneous velocity and temperature measurements, heat removal can be quantified. The time-averaged convective heat flux is calculated as:

$$\phi = \overline{V}\,\overline{T}/U_f T_f \qquad (3)$$

where $\overline{V}$ and $\overline{T}$ are the time-averaged vertical velocity component and temperature. Figure 21 presents the normalized heat flux in the street canyons together with the corresponding flow structure. Significant heat removal is observed in $a_2$, $c_2$ and $d_2$, which is consistent with the flow pattern and temperature profile in the canyon, indicating that these factors dominate the heat removal process.

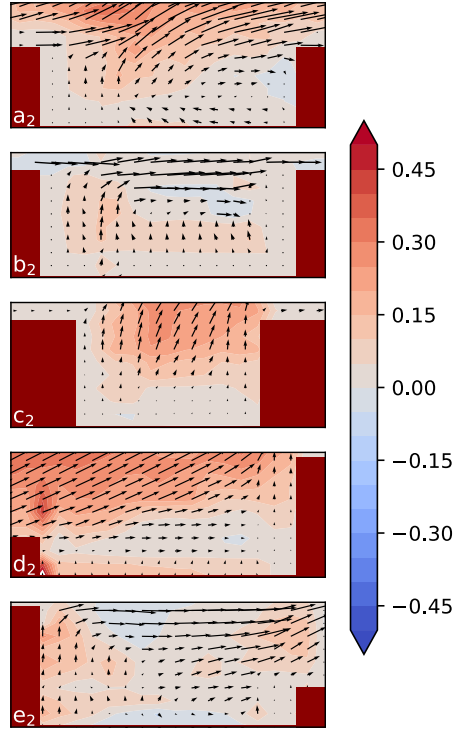

*Figure 21. The time-averaged convective heat flux (ϕ) in all configurations.*

We also calculate the normalized convective heat flux at the roof level as [40]:

$$\emptyset' = \overline{V}(\overline{T_r} - T_f)c_p\rho dA / U_f T_f c_p \rho dA \qquad (3)$$

where $\overline{V}$ and $\overline{T_r}$ are the roof level time-averaged vertical velocity and fluid temperature, $T_f$ the freestream temperature, $c_p$ the heat capacity and $\rho$ density. Figure 22 summarizes the heat flux profiles across the canyon opening, with the position of the narrow canyon (c) indicated by the red dashed line. Case (d) demonstrates the greatest heat removal capability among the tested cases, which can be attributed to its geometric configuration, resulting in the highest ventilation rate. Positive heat removal in cases (a) and (c) occurs throughout the entire canyon opening, while it is only seen on the left side of the canyon (b). Since the fluid temperature on the ground is always higher than the roof temperature (measured by thermal couples), this positive heat removal mechanism is achieved through upward vertical flow. Conversely, Case (e) demonstrates a minimal heat removal effect, with a slight negative value upstream and a positive value downstream. This pattern is determined by the velocity profile within the canyon. Notably, the marked contrast in heat removal performance between canyons with different building heights (Cases (d) and (e)), derived from the same urban model, underscores the significant impact of canyon configuration. It emphasizes the necessity of considering canyon design in the urban planning process, particularly when aiming to mitigate heat-related challenges.

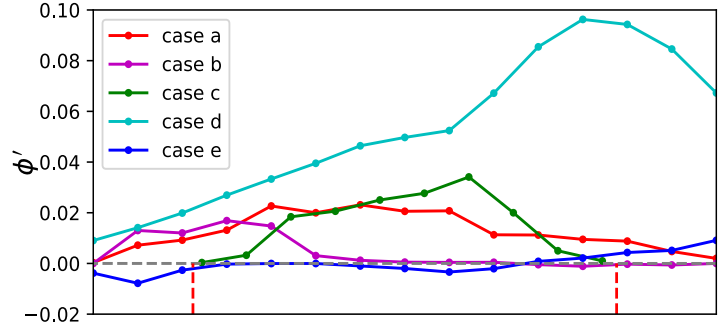

*Figure 22. Normalized heat flux at the canyon opening of all cases over the street canyon width. The red dash line indicates the location of the small canyon in case (c).*

# 7 Further discussions
## 7.1 Impacts of canyon configuration on the canyon flow

Due to the absorption of solar radiation, the surface temperature of the ground and building walls is usually higher than the surrounding air temperature, resulting in a buoyancy-driven boundary layer flow. In the current measurements, the ground is maintained at a constant temperature, and the conductive building models mounted on the heating board allow for convective heat transfer to the flow.

From the above discussion, it can be concluded that the canyon configuration plays a key role in the flow characteristics and heat flux in non-isothermal flow, which is influenced by the interaction between the shear flow and buoyant flow. With the same flow conditions and hence the same shear stress, the total shear force acting on the canyon flow through the canyon opening is directly proportional to the cross-sectional area or, in this case, the width of the canyon in the streamwise direction. Moreover, when considering the same freestream flow condition and surface temperature, the heat removal capacity is primarily determined by the canyon flow, which is influenced by the canyon configuration. As a result, it indicates a direct relationship between the thermal effects and the canyon configuration which is also evident from the temperature profiles depicted in Figure 15. Thermal effects can become more significant if the canyon has a large aspect ratio, which is common in high-density cities and implies narrow canyons. This is because the building walls are also a source of heat, and a large aspect ratio means more heat is transferred to the flow with a unit width of the canyon.

The buoyant flow in a low and wide canyon ($a_2$) with a small aspect ratio of 0.4 and uniform height is strong enough to overcome the suppression of the approaching flow, resulting in a significant enhancement of ventilation and good heat removal performance. However, when only the building height is increased ($b_2$), weakened thermal effects result in less ventilation and poor heat removal. In contrast, a narrow canyon (c$_2$) at the same height has less impact from the shear flow and stronger buoyant flow due to more heat gained from the building models, resulting in good ventilation and heat flux.

The relationship between the shear and buoyant terms is described by the Richardson number, which is commonly used to characterize the non-isothermal flow. The expressions for the Richardson number differ depending on the conditions. Table 2 summarizes the flow properties and Richardson numbers calculated using different parameters in this study. $\Delta T_{s\text{-}f}$ represents the average temperature difference between the heating board surface and the freestream flow, while $\Delta T_{g\text{-}r}$ indicates the temperature difference between the flow over the ground and the flow at the canyon roof level. $U_r$ is the average velocity along the canyon opening and $Ri_{s\text{-}f}$, $Ri_{g\text{-}r}$ are the Richardson numbers calculated based on the freestream and roof level properties, respectively. $\overline{Q'_{iso}}$ and $\overline{Q'_h}$ are the canyon ventilation rates under isothermal and non-isothermal heating conditions, respectively. $\Delta Q'$ represents the enhancement of ventilation, and $\int \phi'$ is the total convective heat flux at the canyon opening.

*Table 2. Summary of the flow properties and parameters in non-isothermal tests*

| Case | $\Delta T_{s\text{-}f}$ | $\Delta T_{g\text{-}r}$ | $U_r$ | $Ri_{s\text{-}f}$ | $Ri_{g\text{-}r}$ | $\overline{Q'_{iso}}$ | $\overline{Q'_h}$ | $\Delta Q'$ | $\int \phi'$ |
|---|---|---|---|---|---|---|---|---|---|
| a | 20 | 2.4 | 0.0034 | 0.8 | 7.691 | -0.35 | 0.91 | 1.26 | 0.155 |
| b | 20 | 2.5 | 0.00587 | 1.1 | 3.584 | -0.10 | -0.07 | 0.03 | 0.061 |
| c | 20 | 2.2 | 0.00422 | 1.1 | 6.102 | -0.08 | 1.28 | 1.36 | 0.136 |
| d | 20 | 3 | 0.0098 | 0.8 | 0.556 | 0.04 | 1.54 | 1.51 | 0.786 |
| e | 20 | 3 | 0.01 | 0.8 | 1.736 | 0.23 | 0.27 | 0.04 | -0.002 |

The effects of thermal conditions on the flow within street canyons are evident from the flow characteristics discussed in the preceding sections, and the significance of these effects can be inferred from the ventilation rates and heat flux. Recent studies have characterized $\Delta T$ using the bulk (average) temperature difference between the freestream flow and the ground in the estimation of the bulk Richardson number, which is widely employed in this field. The thermal force's significance is closely related to the bulk Richardson number under variable heating and flow conditions [25, 26, 30, 41, 42]. However, the bulk Richardson number ($Ri_{s\text{-}f}$) for cases (a, b & c) does not show a clear correlation with the thermal effects in the form of ventilation rate and heat flux, as discussed earlier. For instance, cases b$_2$ and c$_2$ have the same temperature difference, velocity, and model height, resulting in the same Richardson number of 1.1, but the canyon width is different, resulting in a significant difference in ventilation and heat flux.

Indeed, canyons of the same height with the same temperature difference between the heating plate surface and the freestream flow and the same freestream velocity will always have the same Richardson number but may exhibit different heat and fluid removal performances depending on the canyon width. Differences in the average temperature ($\Delta T_{g\text{-}r}$) and velocity ($U_r$) across the height of the canyon, thanks to high-quality measurement velocity and temperature data, are used to calculate the bulk Richardson number ($Ri_{g\text{-}r}$). A good correlation between ($Ri_{g\text{-}r}$) and ventilation/heat flux in canyons with the same height is observed. However, for canyons with variable heights (d & e), the Richardson number does not accurately reflect the thermal impacts in terms of ventilation and heat flux. It has been observed that flow in street canyons with variable height is more complex, with a significant impact on the height difference [25]. Therefore, an effective Richardson number for street canyon flows with complex geometrical configurations and heating conditions is still required [15].

## 7.2 Limitations of the current work

The limitations of this study are primarily due to the narrow range of canyon configurations, heat, and flow conditions investigated. The width, height, and aspect ratio of a canyon can significantly impact the flow characteristics, ventilation, and heat removal within the canyon. To gain a comprehensive understanding of the effects of canyon geometry, it is highly recommended to include more configurations in future investigations. The different non-isothermal flow patterns observed in canyons with varying heights indicate that the heights of the leeward and windward walls can cause significant variations in flow behaviour and thermal effects. A deeper understanding of the flow in such environments can lead to a better definition of the flow characteristic numbers, allowing for a more accurate characterization of the flow instead of relying solely on average height.

Accurately measuring flow properties close to a surface is always a challenging task [43], particularly in non-isothermal flow. The reflection of the laser sheet from the ground in the LIF measurement causes non-uniformly distributed intensity. Furthermore, the presence of thermal boundary layers and heat plumes with strong temperature gradients leads to density variations that are significant enough to generate varying refractive indices. As a result, the optical path of the measurement plane near the heated surface is always changing in LIF measurement, causing measurement errors in both temperature and spatial distribution. Therefore, to minimize these effects, which are only significant close to the surface, only the LIF temperature profiles from 3 mm above the floor are used, and thermocouple measurements are used as validation and correction references. Other limitations in the experiments include the constrained heating capacity of the heating plate, which results in limited heat and flow conditions. The maximum temperature of the plate surface in water is approximately 42°C, resulting in a fluid temperature on the plate of about 28°C. To induce significant thermal effects, the flow is slowed down to 0.03 m/s, resulting in a small magnitude of the freestream Reynolds number ($\sim 10^3$). It is important to note that future experiments with higher heating capacity and a wider range of flow conditions would help to better understand and improve the accuracy of non-isothermal flow measurements.

In investigations involving full-scale field studies, reduced-scale laboratory experiments, and numerical simulations, it is crucial to achieve good similarity of the important non-dimensional numbers characterizing the flow properties. In non-isothermal flows, priority is given to the Richardson number. However, when the similarity of Reynolds number (*Re*) and Richardson number (*Ri*) cannot be satisfied simultaneously, an acceptable treatment in physical and numerical modelling is to relax *Re* by pushing the flow under consideration into a *Re*-independent regime. The similarity and relaxation of characteristic numbers, including the case-dependent and even flow-region-dependent *Re*-independence, have been extensively discussed in the literature [15, 44, 45]. Therefore, to describe the actual flow in a similar urban morphology, it is necessary to pay particular attention to the validity and accuracy of full-scale and lab-scale studies with consideration of *Re*.

We believe that within the limitations, the present study provides new insights into buoyant flow in the street canyon shown by the novel joint observation of the fluid and heat flow in the canyon and the impacts of the canyon's geometrical configurations. This also provides detailed validation data for CFD models having a similar reduced-scale configuration and the Re independence criterion then allows these CFD models to perform full-scale studies by assuming that flows at full scale behave similarly to flows at reduced scale [46].

# 8 Concluding remarks

We report an experimental study focusing on the non-isothermal flow in street canyons using simultaneous PIV/LIF measurements. Our work has made several significant contributions to the understanding of flow and thermal behaviour in street canyons:

- We captured high-resolution velocity and temperature profiles without requiring any distance to the flow, which allowed for a detailed analysis of flow characteristics in different street canyon configurations.
- We investigated the effects of canyon configuration and freestream velocity on canyon flow pattern, roof level ventilation and heat flux, and presented these effects using clearly explained physical processes and analysed parameters.
- Our work demonstrated that the calculated local Richardson number is an effective parameter for characterizing the non-isothermal flow in street canyons.

The experiments are conducted in the ETH Zurich Atmospheric Boundary Layer Water Tunnel at Empa and high temporal and spatial resolution velocity and temperature profiles are captured. Hence, a detailed analysis of the flow characteristics is conducted and summarised with respect to the thermal condition, canyon configuration and freestream velocity magnitude. The changes in flow behaviour in the tested configurations when the ground is heated show the different significance of the thermal condition, which is weakened with increased flow velocity or decreased *Ri*. The local Richardson number calculated using the differences in the average temperature and velocity over the height of the canyon allows a good description of the thermal influence and has a good correlation with the ventilation and heat flux. Quadrant analysis reveals that strong ejections and sweeps events are the dominating components, which are strengthened in heated conditions. Its spatiotemporal plot at the canyon roof level visualizes the periodic-like appearance of the sweeps and ejections and indicates the existence of coherent structures such as plumes. Temperature profiles within the canyon and along the centreline of the canyon are presented, which are related to the flow structure as determined by the model configurations.

The captured information is then used to evaluate the thermal effects in terms of canyon ventilation rate and roof-level opening heat flux. Some significantly enhanced ventilation rates are observed at a freestream velocity of 0.03 m/s in heated conditions. However, it is crucial to acknowledge that this increased ventilation comes at the cost of significantly elevated temperatures within the canyons, which implies a deterioration in thermal comfort conditions within the local microclimate. Our study has unveiled variations in air ventilation and convective heat removal among different canyon configurations under identical heating and flow conditions. For instance, with an increase in canyon height or a decrease in canyon width, the impacts of the approaching flow decrease, and buoyant flow becomes dominant in this chimney-type canyon. With a large canyon width and low canyon height, the shear flow at the canyon opening becomes more important. Canyon height variation could be considered following the principle that the height increases in the direction that the prevailing wind blows. The stepped height concept can help optimize wind capture and guide cooling winds downwards.

While we do not provide explicit design recommendations, our work offers essential insights that can inform future research and guide urban planners, architects, and building engineers in developing more effective strategies to mitigate urban heat challenges. The present work also suggests a further systematic and parametric study, which can lead to a better understanding of the impact of morphology on urban airflow, urban ventilation, and heat removal. The high-quality data obtained from this work can serve as a validation source for numerical studies capable of simulating the full-scale scenario. These simulations hold the potential to yield specific and actionable design recommendations for urban planners, architects, and building engineers, which would be invaluable in devising more effective strategies for addressing urban heat challenges and fostering sustainable urban development.